# Double-discrete solitons in fishnet arrays of optical fibers

Kestutis Staliunas[1] and Boris Malomed[2]

[1]ICREA & Departament de Física i Enginyeria Nuclear, Universitat Politècnica de Catalunya, Colom 11, 08222, Terrassa, Barcelona, Spain

[2]Department of Physical Electronics, School of Electrical Engineering, Faculty of Engineering, Tel Aviv University, Tel Aviv 69978, Israel

**Abstract:** We demonstrate that crossed arrays of optical fibers support the double-discrete linear and nonlinear propagation of light beams, in which not only the transverse coordinate (the fiber's number) is discrete, but also the longitudinal (propagation) coordinate, i.e., the number of the fiber-crossing site, is effectively discrete too. In the linear limit, this transmission regime features double-discrete self-collimation. The nonlinear fishnet arrays with both focusing and defocusing nonlinearities give rise to *double-discrete* spatial solitons. Solitons bifurcating from two different branches of the linear dispersion relation feature strong interactions and form composite states. In the continuum limit, the model of the nonlinear fishnet reduces to a system of coupled-mode equations similar to those describing Bragg gratings, but without the cross-phase-modulation terms.

## I. Introduction

Arrays of coupled optical fibers or waveguides, both uniform and modulated ones, are discrete media in which one can modify ("manage") the effective dispersion and diffraction, and thus control the linear and nonlinear light propagation [1]. In particular, the "diffraction-management" schemes make it possible to reverse the sign of the effective diffraction [2]. In nonlinear media, the use of these techniques helps to create gap solitons, as predicted theoretically [3] and realized experimentally [4]. Further, the diffraction can be set to zero in zigzag arrays [5], leading to an effect similar to the self-collimation in photonic crystals with eliminated diffraction [6,7]. Nontrivial linear and nonlinear effects were also predicted and observed in periodically snaking [8] and antiphase-snaking [9,10] arrays, in arrays with more sophisticated periodic modulations [11], as well as in "blinking" and "shaking" [12] Bragg gratings. The linear and nonlinear propagation effects for light in fiber arrays have their counterparts in Bose-Einstein condensates (BECs) loaded into modulated potentials: the snaking or zigzagging fiber arrays are analogous to "shaking" periodic lattices [15,16]. In particular, various effects for solitons can be induced by means of these techniques, which place them into the general class of the "soliton management" methods [17].

Here we propose a novel but quite simple configuration of fiber arrays, patterned as a fishnet, i.e., a pair of crossed arrays. We demonstrate that this configuration allows a rich variety of light-propagation phenomena in linear and nonlinear regimes. The diffraction may be effectively reduced to zero in this setting, as well as made negative. Moreover, several distinct dispersion curves coexist in the system. This allows the self-collimating behavior (which are known in other configurations too), as well as periodic beatings of co-propagating collimated modes – a regime which was not reported before. The configuration also gives rise to special features of the nonlinear propagation: double-discrete solitons, similar to those of the bandgap type, and coexistence of different species of solitons bifurcating from different branches of the dispersion curves – again an effect not reported before in such settings, to the best of our knowledge. Furthermore, solitons originating from different branches may form stable composite states. This variety of linear and nonlinear regimes, combined with the relative simplicity of the structure, and potential ease of its fabrication, is a motivation of the present analysis.

It is relevant to mention that the double-discrete solitons somewhat resemble those found in models of integrable automata [18]. However, the actual form of those models is very different, and they can hardly be implemented in optics.

The paper is organized as follows: in Section II, we introduce the proposed scheme and define a dynamical map which furnishes its description. In Section III we solve the map in the linear regime and produce the respective dispersion curves, discussing different combinations of such curves. We also demonstrate peculiar behavior of light, following from the dispersion curves, with the help of direct numerical integration of the linear map. Double-discrete solitons in the system with the cubic nonlinearity are constructed in a numerical form in Section IV. We analyze the soliton dynamics in terms of the map, with emphasis on the coexistence and collisions of the solitons bifurcating from different branches of the dispersion relation. Next, we study moving and colliding solitons in Sec. V. Finally, we get back to analytics, deriving a continuous limit of the map, in the form of coupled-mode equations, in Section VI. The paper is concluded by Section VII.

**II. The fishnet system**

The configuration of the fiber arrays is sketched in Fig. 1(a). It is composed of two arrays (layers) of parallel-coupled fibers, each layer built as proposed in Ref. [19]. The two arrays form an angle, being linearly coupled at crossing points, as indicated in the top view of the system in Fig. 1(b). This figure illustrates what is expected in the linear regime: the light can propagate without diffraction, or with very weak diffraction, roughly along diagonals of rhombuses of the fishnet structure, under conditions which are produced below.

It is relevant to note that a double-discrete fiber network was recently realized experimentally and analyzed theoretically in Ref. [20]. The linear regime, investigated in that work, features unusual linear evolution in the presence of controllable fiber loss, such as subexponential decay and formation of fractal patterns. Furthermore, this system was also used to experimentally implement the double-discrete system subject to the condition of the *PT* symmetry between gain and loss elements [21].

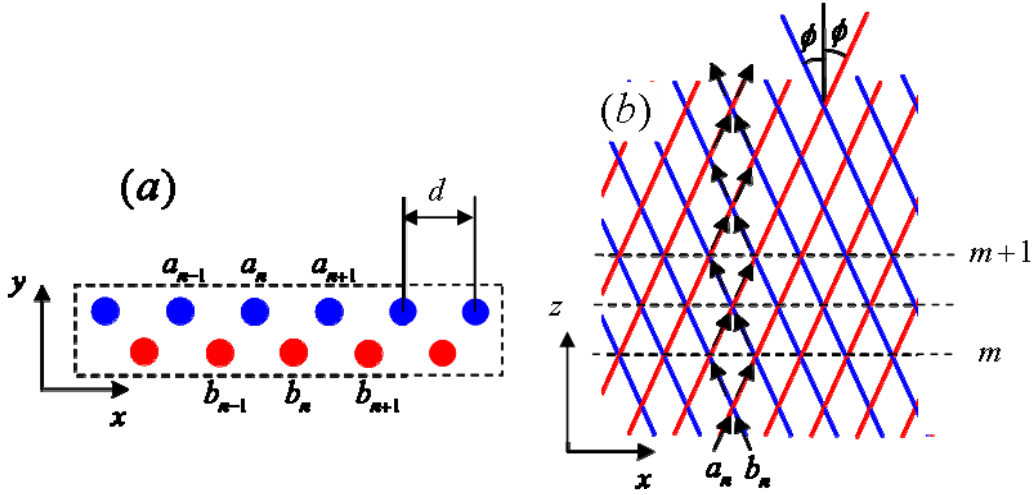

*Fig. 1.* (Color online) The scheme of the fiber arrays: a) the front view [a cross section in the (x,y) plane; the horizontal shift between the fibers in top and bottom arrays vanishes at integer and half-integer values of z corresponding to the intersection between the two arrays]; b) the top view in the (x,z) plane. Arrows schematically show the expected diffraction-free propagation of light.

*The propagation map.* The full map describing the propagation of light in the fishnet system is built of several consecutively applied propagation and interaction operators, acting within one period. Each operator is presented separately below.

*The linear propagation.* First we consider the linear transmission of the field amplitudes in each array, $a_n$ and $b_n$, from a given node to the next one, which is driven by the tunnel coupling between adjacent fibers in each array:

$$da_n/dz = ic(e^{+i\alpha}a_{n-1} + e^{-i\alpha}a_{n+1}), \tag{1a}$$

$$db_n/dz = ic(e^{-i\alpha}b_{n-1} + e^{+i\alpha}b_{n+1}). \tag{1b}$$

Here *c* is the coupling constant, and $\alpha$ is determined by the tilt of the arrays:

$$\alpha = 2\pi d \lambda^{-1} \sin\phi, \tag{1c}$$

where, as indicated in Fig. 1(b), *d* is the distance between the fibers, $\pm\phi$ are the tilt angles of the two arrays, and $\lambda$ is the wavelength of light. These equations describe the slow evolution (along z) of the amplitudes of waveguide modes, while the rapidly oscillating factor, $\exp(iK_0 z)$, with carrier wavenumber, $K_0$, is eliminated.

The Fourier transform of Eqs. (1), with $a_n \equiv \int_{-\pi}^{+\pi} a_k \exp(ikn) dk$, $b_n \equiv \int_{-\pi}^{+\pi} b_k \exp(ikn) dk$, takes the form:

$$da_k/dz = 2ica_k \cos(k-\alpha), \tag{2a}$$

$$db_k/dz = 2icb_k \cos(k+\alpha). \tag{2b}$$

Equations (2) can be easily integrated, which allows us to define an explicit map for the transmission between nodes of the grid:

$$a'_k = a_k \exp(2i\gamma \cos(k-\alpha)), \tag{3a}$$

$$b'_k = b_k \exp(2i\gamma \cos(k+\alpha)). \tag{3b}$$

Here $\gamma \equiv cl$, where $l = d/\sin(2\phi)$ is the length of the grid's cell, is the effective strength of the coupling between adjacent nodes of the present network.

Equations (3) can be rewritten in a more convenient matrix form, $A'_k = \hat{L} A_k$, with vectors $A_k \equiv (a_k, b_k)$ evolving under the action of the linear propagation operator:

$$\hat{L} = \begin{pmatrix} \exp(2i\gamma \cos(k-\alpha)) & 0 \\ 0 & \exp(2i\gamma \cos(k+\alpha)) \end{pmatrix}. \tag{4}$$

Note that all relevant values of phase shift $\alpha$, which is defined by Eq. (1c), are accessible in the experiment. In particular, in the regime of $d \gg \lambda$, one can easily scan the whole interval $-\pi < \alpha < +\pi$ of $\alpha$, tuning the tilt angle $\phi$ by a small amount.

*The inter-array coupling.* At the crossing nodes, the two arrays are coupled to each other, which is described by operator

$$\hat{M} = \begin{pmatrix} \cos(\mu) & i\sin(\mu) \\ i\sin(\mu) & \cos(\mu) \end{pmatrix}, \tag{5}$$

which is identical in the *n*- and *k*-spaces. Here $\mu$ is the coefficient of the coupling of two fibers, belonging to the different layers, at their intersection points.

*Re-indexing of the fibers.* With the reference frame fixed as shown in Fig. 1(b), the fibers are to be re-indexed after each crossing. To this end, the discrete coordinates of the two arrays are shifted to the right and to the left respectively, which, in the *k*-space, is realized by the action of the following operator:

$$\hat{S}_A = \begin{pmatrix} \exp(-ik) & 0 \\ 0 & 1 \end{pmatrix}; \qquad \hat{S}_B = \begin{pmatrix} 1 & 0 \\ 0 & \exp(+ik) \end{pmatrix}. \tag{6}$$

*The full-period linear propagation.* The full map describing the linear evolution of the field over one period is thus given by $\hat{P} = \hat{S}_B \hat{M} \hat{L} \hat{S}_A \hat{M} \hat{L}$, where the operators are applied consecutively from right to left. Note that the above-mentioned re-indexing of the arrays is applied after the subsequent crossings of the fibers (i.e., twice per full period).

*Nonlinearity*. The nonlinearity, if it is continuously distributed, should be considered together with the propagation in each lattice cell, as described by Eq. (1), i.e., by means of the corresponding nonlinear modification of the propagation operator, $\hat{L}$. To make the model tractable, we here assume that the nonlinearity is concentrated at one point in each period, as in the so-called split-step system [22] (i.e., it may be represented by a highly nonlinear segment inserted into each fiber span). Thus, the nonlinearity acts in the coordinate (*n*) domain, being represented by the following self-phase-modulation (SPM) operator:

$$\hat{N} = \begin{pmatrix} \exp(i\delta|A_n|^2) & 0 \\ 0 & \exp(i\delta|B_n|^2) \end{pmatrix}. \tag{7}$$

The nonlinearity coefficient, $\delta$, is positive (negative) for the focusing (defocusing) nonlinearity. Adding operator (7), the full map reads explicitly as $\hat{P} = \hat{N}\hat{S}_B\hat{M}\hat{L}\hat{S}_A\hat{M}\hat{L}$.

**III. The linear system**

The linear propagation is determined by the dispersion relation, which is obtained by calculating eigenvalues of the propagation operator, $\hat{P}$. Properties of nonlinear modes (solitons) also strongly depend on the form of the linear dispersion.

As the propagation is conservative, the absolute value of the eigenvalues is unity, which corresponds to linear modes in the form of $a_{k,m+1} = a_{k,m}\exp(i\beta)$, $b_{k,m+1} = b_{k,m}\exp(i\beta)$, where $\exp(i\beta)$ are the Floquet multipliers. Phase $\beta$ of the multipliers, which depends on *k*, determines the dispersion relation (in continuous models, the dispersion relation is represented by the dependence of the propagation constant on *k*). Figure 2 shows a collection of dispersion curves for different combinations of parameters $\alpha, \lambda, \mu$, which represent the tilt of the arrays and strengths of the in- and inter-array coupling.

A typical result is the appearance of bandgaps around cross-point of the dispersion curves of the two uncoupled tilted arrays. The latter curves, shown by dashed lines in Fig.2, are produced by operators (4) and (6) for the uncoupled arrays:

$$\beta_{1,2} = \pm k + 2\gamma \cos(k \pm \alpha), \tag{8}$$

and is. The $2\pi$-periodicity of the propagation wavenumber in Fig. 2 is a consequence of the discreteness of the system.

The presence of the bandgap is a necessary condition for the existence of gap solitons in nonlinear systems (we consider only bright solitons in this work). As Fig. 2 shows, all possible combinations of curvatures of the dispersion curves may occur around the bandgap. In particular, the interplay of the positive-positive combination in Fig. 2(a) with the focusing nonlinearity may result in coexistence of two families of bandgap solitons, bifurcating from each dispersive branch [23]. The negative-negative combination in Fig. 2(b) results in two families of solitons for the defocusing nonlinearity, and the positive-negative combination in Fig 2(e) gives rise to the "classical" bandgap solitons for either sign of the nonlinearity.

An important result is the appearance of flat segments in the dispersion curves for particular combinations of the coefficients, see Fig. 2(d). This fact, first of all, implies the onset of the subdiffraction or self-collimation, in terms of the linear map. In the nonlinear system, it suggests searching for very narrow (*subdiffractive*) solitons for either sign of the nonlinearity. In other words, in the subdiffractive regime a very weak nonlinearity is sufficient to compensate the very weak subdiffraction, and thus to form solitons. Similar subdiffractive solitons were reported in shaking [14] and blinking [15] lattices. Panels (c) and (f) in Fig. 2 show more complex shapes of the dispersion.

In the diagrams displayed in Fig. 2, the bandgaps occur only around $k=0$. A careful numerical analysis demonstrates that the interaction between the two arrays, even if it is strong, does not open extra gaps at points $k=\pm\pi$, where different dispersion curves cross too. The intersection between different dispersion branches is possible if the corresponding Bloch modes remain mutually orthogonal. In this connection, it is relevant to mention that, in the general case, gap solitons cannot be created by the nonlinearity if the bandgap is absent in the linear spectrum [24], except for a special case when the gap actually exists, but its width is zero, as one of the dispersion lines is asymptotically horizontal [25].

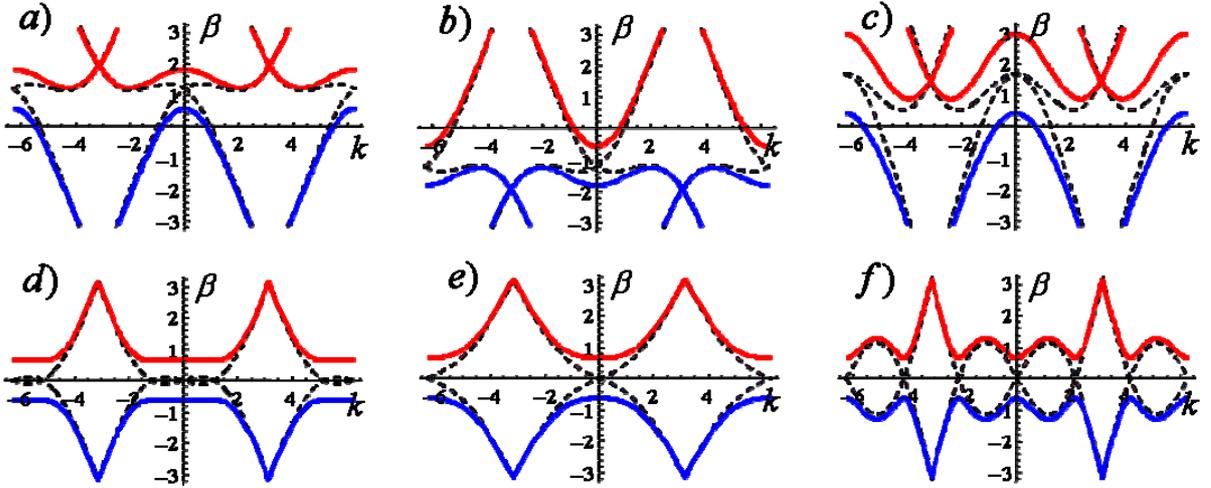

*Fig.2. (Color online) Dispersion curves of the coupled arrays. Represented is the phase of the Floquet multiplier ($\beta$) versus the transverse wavenumber, k. To visualize the opening of the bandgaps, the dispersion of uncoupled arrays is shown by dashed lines. Several typical configurations are shown: a) both dispersion lines negatively curved (the normal dispersion), with different absolute values of the curvatures, $\alpha=0.1\pi, \gamma=0.1\pi, \mu=0.1\pi$; b) both lines positively curved, $\alpha=0.9\pi, \gamma=0.1\pi, \mu=0.1\pi$; c) both lines with approximately the same (normal) curvature, $\alpha=0.2\pi, \gamma=0.15\pi, \mu=0.15\pi$; d) a broad flat segment, implying the self-collimation, $\alpha=0.5\pi, \gamma=0.1\pi, \mu=0.1\pi$; e) a "classical" bandgap structure, $\alpha=0.5\pi, \gamma=0.1\pi, \mu=0.05\pi$; f) a multi-bandgap structure, $\alpha=0.5\pi, \gamma=0.1\pi, \mu=0.2\pi$.*

To test the linear dispersion relations, we have performed numerical simulations of the linear fishnet model. Two characteristic examples of the so generated evolution are displayed in Fig. 3 (the usual discrete diffraction) and Fig. 4 (the self-collimation). The observed behavior corroborates what could be expected from the calculated dispersion curves.

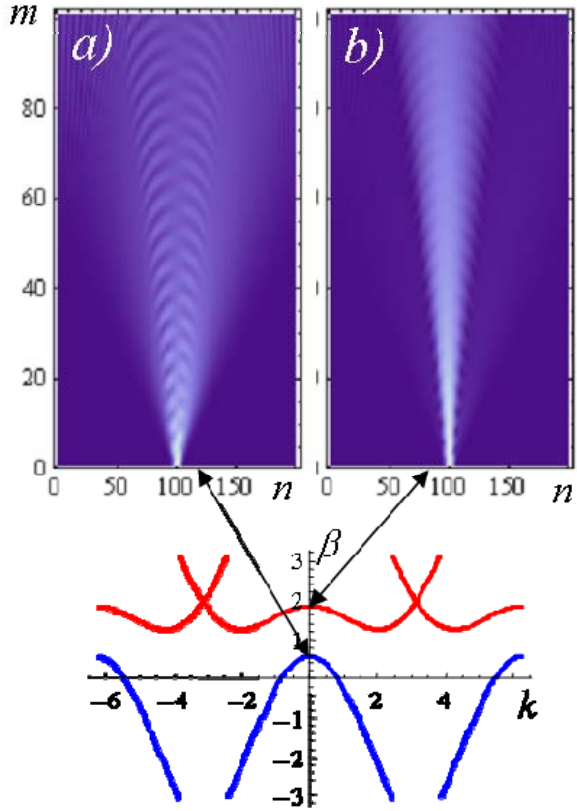

*Fig.3.* (Color online) Examples of the linear propagation in the case when the two dispersion branches have negative curvatures with different absolute values. The upper branch is excited when the field components are injected into the upper and bottom arrays with the same phase (the corresponding eigenvector is (1,1)), and the lower branch – when the components are π-out-of-phase, with the corresponding eigenvector (1,-1). Parameters are $\alpha = 0.1\pi, \gamma = 0.1\pi, \mu = 0.1\pi$.

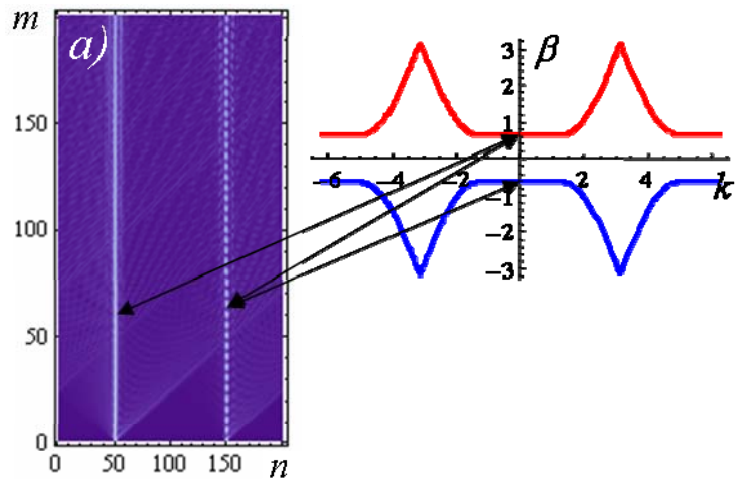

*Fig.4.* (Color online) Examples of the linear propagation in the self-collimation regime. Two narrow beams were injected in the same integration domain. The left beam is the envelope of the Bloch mode with eigenvector (1,1), while the right one has (1,0), therefore it splits into two Bloch modes with eigenvectors (1,1) and (1,-1). The split modes have different propagation constants, which results in periodic beatings (between the upper and lower arrays) exhibited by the right beam. Parameters are $\alpha = 0.5\pi, \gamma = 0.1\pi, \mu = 0.1\pi$.

**IV. Numerical results for solitons in the nonlinear system**

Proceeding to the nonlinear system, in Fig. 5 we present numerical simulations for solitons in the case (arguably, most interesting one), when the two branches of the dispersion curves feature the curvature of the same sign at $k = 0$. In fact, solitons can be found for all the typical shapes of the dispersion curves presented in Fig. 2.

As could be expected, the solitons bifurcating from two different dispersion branches coexist for the same set of parameters. The linear dispersion implies different diffraction coefficients for the Bloch waves associated with the upper and lower branches (the diffraction coefficient is proportional to the curvature of spatial-dispersion line). Therefore, the corresponding solitons with equal amplitudes are expected to have different widths.

An unexpected and quite interesting finding is that, apart from the "pure" soliton states, composite ones, which lock together pulses emerging from the different dispersion branches, exist too. These mutually trapped composite modes propagate at a common "velocity" (actually, the spatial-domain tilt) and feature periodic dynamics in the form of beating, i.e., oscillations in $z$. The oscillation period depends on the separation between the branches, and also on nonlinear phase shifts for both soliton components. Qualitatively similar composite states are known too in continuous systems (in particular, mutually trapped states of fiber solitons with different polarizations [26]).

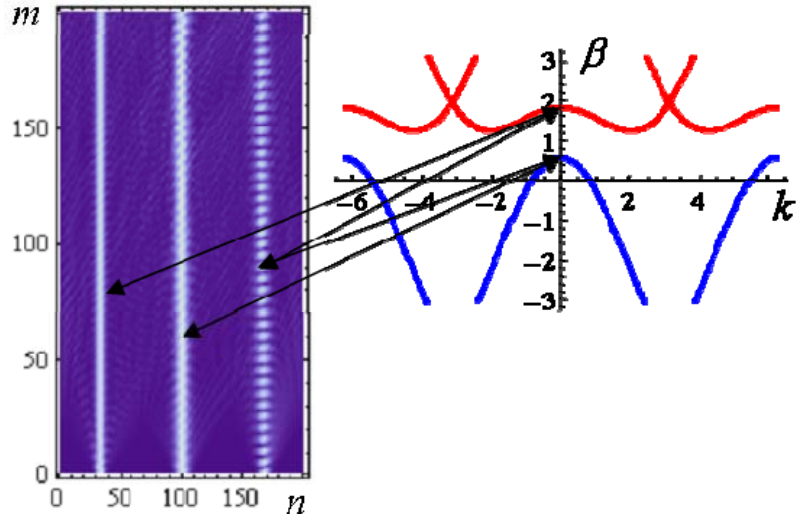

*Fig.5. (Color online) The solitons bifurcating from the upper branch [the corresponding eigenvector is (1,1)], from the lower one corresponding to eigenvector (1,-1), and a composite oscillating mode, generated by the injection of the field configured as (1,0), solely into the top layer. It is seen that widths of the solitons are different, as the two branches have different curvatures at $k_0 = 0$. Parameters are $\alpha = 0.1\pi$, $\gamma = 0.1\pi$, $\mu = 0.1\pi$, $\delta = 0.2$.*

Stability of both types of solitons, bifurcating from the two dispersion branches, requires a special consideration. Numerically calculating propagation parameters for the solitons bifurcated from the two branches, we have found that their values are close to those at the corresponding bifurcation points. The calculated envelopes of the solitons are displayed in Fig. 6. Differently from the above other calculations with periodic boundary conditions, see Figs. 3-5, here we use absorbing boundaries to eliminate the outgoing radiation. This ingredient of the numerical scheme is important, as the solitons might be destabilized by radiation artificially re-entering through the boundaries.

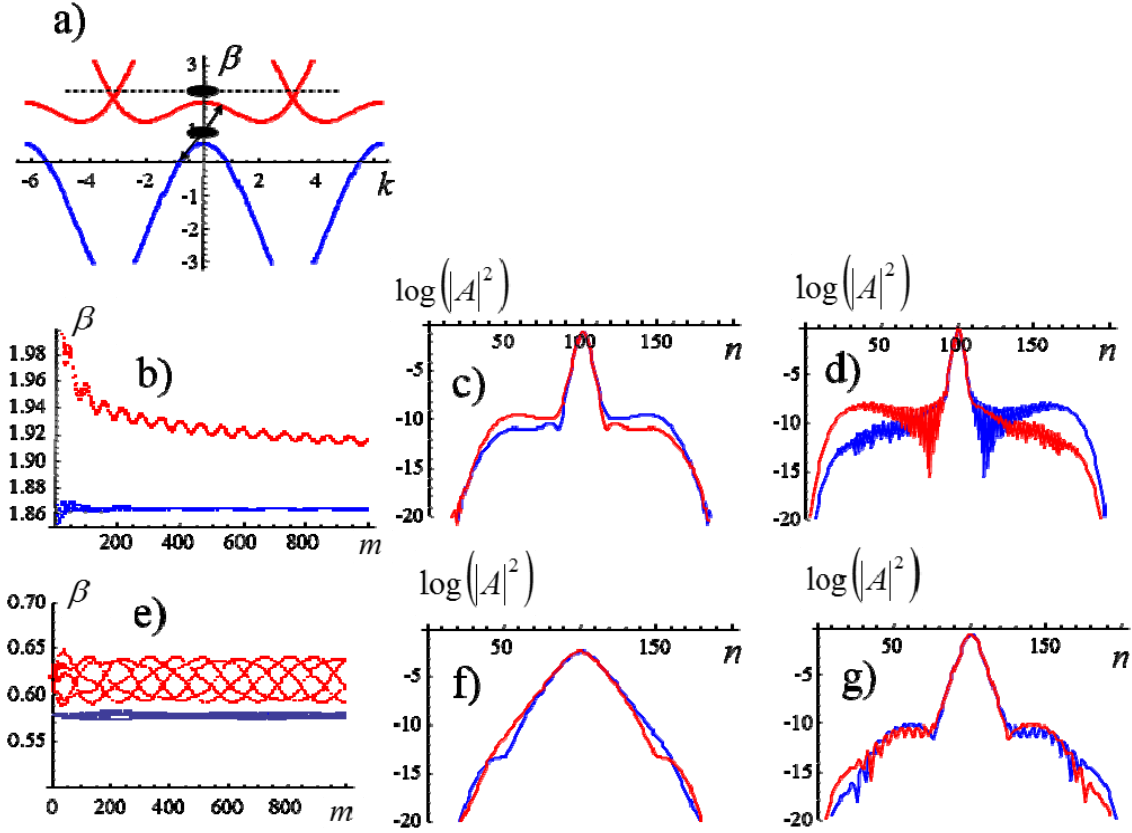

*Fig. 6. (Color online) The diagram illustrating the instability mechanisms of the solitons bifurcated from the two branches of the dispersion curves. (a) The black spots designate the spectral location of the solitons with respect to the dispersion curves. (b) The evolution of the soliton's propagation constant, $\beta$ for the intensely radiating and practically stationary solitons (the top red and bottom blue plots, respectively) bifurcated from the upper dispersion branch solitons. (c) Examples of effectively unstable and practically stable bandgap solitons, bifurcated from the lower dispersion curve, and interacting with the radiation modes through the FWM process, as illustrated by arrows in panel (a). Shapes, shown on the linear-logarithmic scale, of the two components (pertaining to the two arrays in the fishnet) of solitons for different propagation constants: (c) $\beta = 1.865$, (d) $\beta = 1.92 \pm 0.005$, (f) $\beta = 0.58$, (g) $\beta = 0.62 \pm 0.03$.*

The solitons bifurcated from the upper branch generally emit the Cherenkov radiation, due to the resonant coupling to the continuum modes. The dashed horizontal line in Fig. 6(a) indicates the resonance of the soliton with the continuum modes. However the radiation intensity strongly depends on the width of the soliton's spectrum.

The solitons which are spatially narrower, having a larger amplitude (with a larger propagation constant, $\beta$) radiate more. Due to the Cherenkov emission, the soliton loses its energy, becomes broader, hence its radiation losses decreases. Strictly speaking, these solitons are unstable (in fact, they do not exist as rigorously defined stationary solutions), slowly drifting back to the bifurcation point. However, the radiation loss rate, which is proportional to the squared overlap integral of the soliton's spectrum with the continuum modes, becomes exponentially small in the course of the evolution. Figure 6(b) shows examples of a relatively strongly radiating soliton, and of a quasistationary one, both bifurcated from the upper dispersion branch.

The soliton bifurcated from the bottom branch, whose propagation constant belongs to the true bandgap, can also radiate, and thus can display the instability. However, this radiation is generated by the four-wave-mixing (FWM) interaction [27], rather than representing the Cherenkov resonance with the continuum (similar to the possible mechanism of the FWM-mediated subharmonic decay of intrinsic localized modes in nonlinear lattices [28]). The pair of arrows connecting the bandgap (lower) soliton in Fig. 6(a) with the two dispersion branches designate the scheme of the FWM emission, where the soliton loses its photons in pairs, due to the mixing with the radiation modes belonging to the upper and to the bottom bands (note that the oppositely directed arrows have equal lengths, which should ensure the energy and momentum conservation in the course of the FWM process). Examples of a bandgap soliton destabilized by the FWM interaction with the radiation modes, and of a virtually stable one, are shown in Fig. 6(e).

In the standard model with symmetric bands [29], gap solitons are stable in approximately the bottom half of the bandgap [30]. In the case of asymmetric bands, the stability area may be different (in our case, it is smaller, as a simple geometric consideration prompts). Nevertheless, the solitons are stable sufficiently close to the bifurcation point, see examples of completely stable solitons in Figs. 6(c,f). Farther from the bifurcation point, they develop oscillatory tails and a small uncertainty in the value of the propagation constant, as Figs. 6(d,g) demonstrate.

**V. Moving solitons.**

The solitons may be imparted the above-mentioned tilt (i.e., the spatial-domain counterpart of the velocity), multiplying the input by $\exp(ik_0 n)$, with transverse kick $k_0$. For fixed $k_0$ the solitons originating from different dispersion branches propagate at different angles, as the slopes of the dispersion curves are different. Examples of the moving solitons corresponding to the different branches, as well as of the composite soliton, are displayed in Fig.7.

We studied collisions between tilted solitons. As seen in Fig. 8, the collisions seem generally inelastic, which is not surprising, as the system is not close to any integrable limit. Actually, collisions between the solitons originating from the lower dispersion branch seem more quasi-elastic (perhaps because this branch is closer to an ideal parabola). Least elastic is the collision between solitons belonging to the different branches: as a result, the solitons tend to merge into the above-mentioned composite states, on the contrary to a "naïve" expectation that the interaction would be weakest between solitons originating from different linear modes.

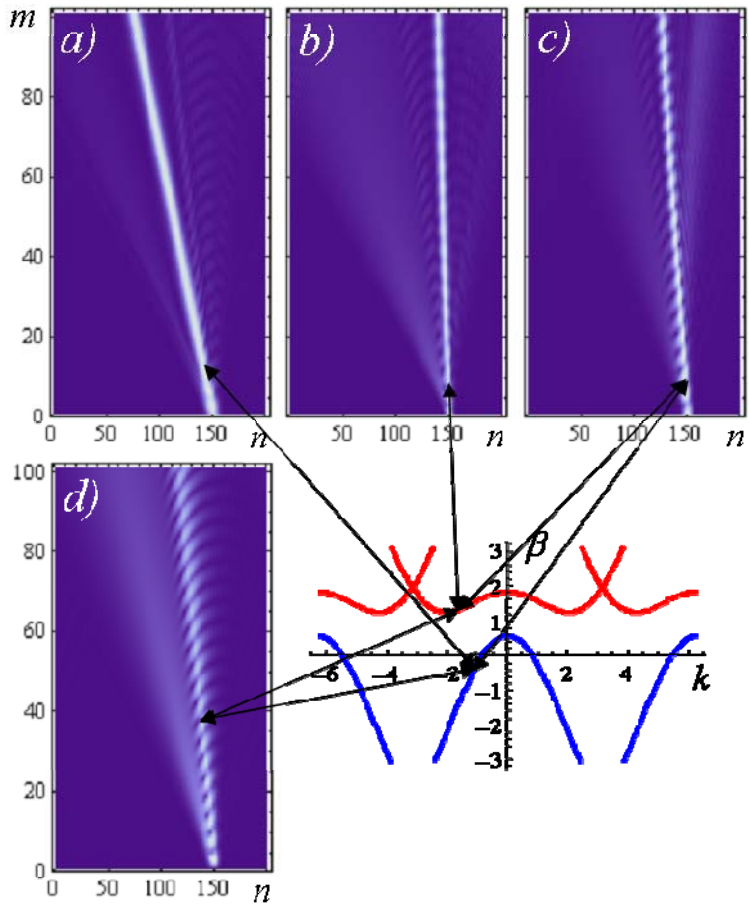

*Fig. 7.* (Color online) Solitons produced by obliquely injected beams, in this case with transverse kick $k_0 = -0.5\pi$: a) the soliton of the (1,-1) type (a broader one belonging to the lower branch); b) the soliton of the (1,1) type (a narrower beam, pertaining to the upper branch); c) and d) composite states, obtained from (1,0) and (0.1) injected beams, respectively. Parameters are as in Fig.5.

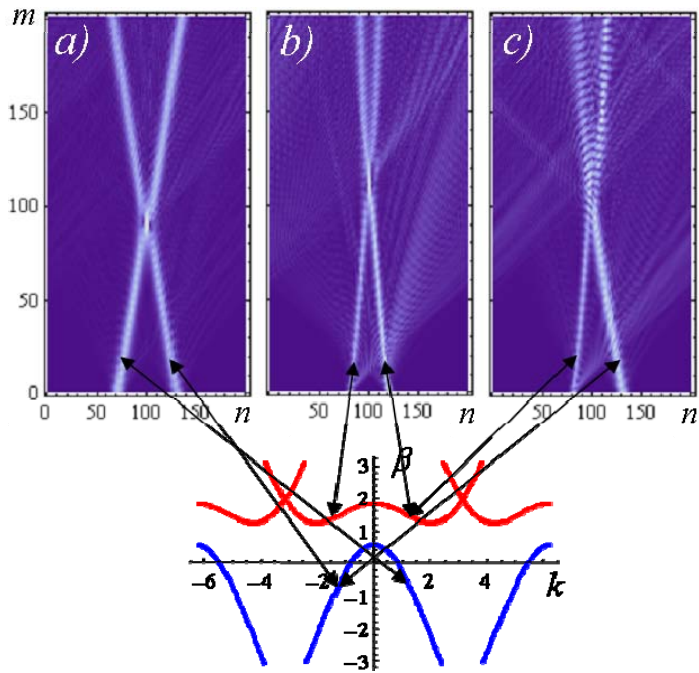

*Fig. 8.* (Color online) Generic examples of collisions between oblique solitons belonging to the same or different branches, kicked with $k_0 = \pm 0.5\pi$: a) both belong to branch (1,-1); b) both belong to branch (1,1); c) the solitons belong to different branches (1,-1) and (1,1). Parameters are as in Fig.5.

**VI. The continuum limit**

In the long-wave limit, when the transverse wavenumber $k$ becomes vanishingly small, a straightforward expansion of the dispersion relation for the uncoupled arrayed layers, given by Eq. (8), yields

$$\beta_{1,2} = 2\gamma\cos(\alpha)\left(1 - \frac{k^2}{2}\right) \pm k(1 - 2\gamma\sin(\alpha)) + ... \quad . \tag{9}$$

The corresponding system of continuous *coupled-mode equations* for waves $u(x,z)$ and $v(x,z)$ in the two layers, where $x$ is the continual counterpart of $n$, can be then derived, replacing $k$ by $i\partial/\partial x$, absorbing the zero-order term in expansion (9), $2\gamma\cos(\alpha)$, into a shift of the carrier propagation constant, and expanding the coupling and SPM operators, (5) and (7), for the cases when both the coupling and SPM nonlinearity are weak (to be in the balance with the weak spatial dispersion corresponding to the long-wave limit):

$$i\frac{\partial u}{\partial z} + i(1 - 2\gamma\sin(\alpha))\frac{\partial u}{\partial x} + \gamma\cos(\alpha)\frac{\partial^2 u}{\partial x^2} + \delta|u|^2 u + \mu v = 0, \tag{10a}$$

$$i\frac{\partial v}{\partial z} - i(1 - 2\gamma\sin(\alpha))\frac{\partial u}{\partial x} + \gamma\cos(\alpha)\frac{\partial^2 u}{\partial x^2} + \delta|v|^2 v + \mu u = 0, \tag{10b}$$

The linearization of Eqs. (10) yields the continuum-limit form of the dispersion relation for the fishnet medium: $\beta_{1,2} = 2\gamma\cos(\alpha)\left(1 - \frac{k^2}{2}\right) \pm \sqrt{k^2(1 - 2\gamma\sin(\alpha))^2 + \mu^2}$, which provides a good approximation for the generic dispersion curves of the double-discrete system, such as those displayed above in Fig. 2(e).

If characteristic wavenumbers in Eqs. (9) and (10) are small in comparison with the inter-layer coupling constant $\mu$, the second-derivative terms (intrinsic dispersion) may be dropped in Eqs. (10), in comparison with the effective dispersion induced by the linear coupling, cf. Ref. [31]. Then, after an obvious rescaling, the simplified version of Eqs. (10) equations is cast into a parameter-free form:

$$\begin{aligned} i\frac{\partial u}{\partial z} + i\frac{\partial u}{\partial x} + |u|^2 u + v &= 0, \\ i\frac{\partial v}{\partial z} - i\frac{\partial u}{\partial x} + |v|^2 v + u &= 0. \end{aligned} \tag{11}$$

Equations (11) are equivalent to the standard system of coupled-mode equations for the co-propagation of two waves in Bragg gratings, provided that the nonlinear cross-phase-modulation (XPM) terms are absent. Indeed, the structure of our model implies that the XPM, which acts solely at the crossing points, is negligible in comparison with the intrinsic self-phase modulation acting in each array. Thus, the continuum limit of the fishnet systems offers an interesting example of the coupled-mode system with the solely-SPM type of the nonlinearity. In the usual fiber-Bragg gratings [29], as well as in spatial gratings [32], the coupled-mode equations contain XPM terms, which are stronger than their SPM counterparts by a factor of 2.

Finally, it is well known that Eq. (11) has a family of exact gap-soliton solutions, both standing and walking ones [29], roughly half of which is stable [30], as mentioned above. Those exact solutions provide for the continuum counterpart of the double-discrete bandgap solitons, such the one shown above in Fig. 6(f).

## VII. Conclusions

We have proposed a fishnet-shaped optical-waveguiding system, which implements the double-discrete linear and nonlinear transmission of optical beams. The linear regime readily features self-collimation and periodic beatings of overlapping collimated beams belonging to two different branches of the dispersion relation. The nonlinear system gives rise to stable double-discrete solitons, including composite ones, formed by the solitary modes bifurcating from the different dispersion branches. The continuum limit of the setting reduces to the system of coupled-mode equations, without the XPM terms.

A challenging extension of the present setting would be to consider not only the "forward" but also "backward" propagation throughout the fishnet (reflections). This may happen when the crossing angles are large. The analysis of this case should be different (iterations should be done in time, rather than by marching along the grid). This generalization will be considered elsewhere. It may also be interesting to consider an extension of the present system including losses and gain.

## Acknowledgments

K.S acknowledge financial support by Spanish Ministerio de Educación y Ciencia and European FEDER (project FIS2011-29734-C02-01) and Generalitat de Catalunya (2009 SGR 1168). B.A.M. appreciates hospitality of ICFO-Institut de Ciencies Fotoniques (Barcelona, Spain).